\documentclass[preprint]{aastex}
\usepackage{color}
\usepackage{natbib}
\usepackage{multirow}

\newcommand{\Bv}{\mathbf{v}}
\newcommand{\BB}{\mathbf{B}}
\newcommand{\Bj}{\mathbf{j}}
\newcommand{\BA}{\mathbf{A}}

\newcommand{\dAdt}{\frac{\partial\mathbf{A}}{\partial t}}

\shorttitle{How Magnetic Topology can Kill a Solar Eruption}
\shortauthors{Chintzoglou et al.}

\begin{document}

\title{Magnetic Flux Rope Shredding by a Hyperbolic Flux Tube: The Detrimental Effects of Magnetic Topology on Solar Eruptions}

\author{Georgios Chintzoglou}
\affil{Lockheed Martin Solar and Astrophysics Laboratory,
	                3176 Porter Dr, Palo Alto, CA 94304, USA}

\affil{University Corporation for Atmospheric Research,
       Boulder, CO 80307-3000,USA}
\email{gchintzo@lmsal.com}

\and

\author{Angelos Vourlidas\altaffilmark{1}}
\affil{The Johns Hopkins University Applied Physics Laboratory, Laurel, MD 20723, USA}

\and

\author{Antonia Savcheva   and   Svetlin Tassev}
\affil{Harvard-Smithsonian Center for Astrophysics, Cambridge, MA 02138, USA}

\and
\author{Samuel Tun Beltran   and   Guillermo Stenborg}
\affil{Space Science Division, Naval Research Laboratory, Washington, DC 20375, USA}

\altaffiltext{1}{Also at IAASARS, National Observatory of Athens, GR-15236, Penteli, Greece}

\begin{abstract}

We present the analysis of an unusual failed eruption captured in high cadence and in many wavelengths during the observing campaign in support of the \emph{VAULT2.0} sounding rocket launch. The refurbished Very high Angular resolution Ultraviolet Telescope (\textit{VAULT2.0}) is a Ly$\alpha$ ($\lambda$ 1216\,\AA) spectroheliograph launched on September 30, 2014. 
The campaign targeted active region NOAA AR 12172 and was closely coordinated with the \textsl{Hinode\/} and \textsl{IRIS\/} missions and several ground-based observatories (NSO/IBIS, SOLIS, and BBSO). A filament eruption accompanied by a low level flaring event (at the GOES C-class level) occurred around the \emph{VAULT2.0} launch.
No Coronal Mass Ejection (CME) was observed. The eruption and its source region, however, were recorded by the campaign instruments in many atmospheric heights ranging from the photosphere to the corona in high cadence and spatial resolution. This is a rare occasion which enables us to perform a comprehensive investigation on a failed eruption. We find that a rising Magnetic Flux Rope-like (MFR) structure was destroyed during its interaction with the ambient magnetic field creating downflows of cool plasma and diffuse hot coronal structures reminiscent of ``cusps''. We employ magnetofrictional simulations to show that the magnetic topology of the ambient field is responsible for the destruction of the MFR. Our unique observations suggest that the magnetic topology of the corona is a key ingredient for a successful eruption. 
\end{abstract}

\section{Introduction}
The magnetic field drives the dynamics of the solar atmosphere, from the highly structured extreme ultraviolet (EUV) corona to the spectacular coronal mass ejections (CMEs). The quest for the cause of eruptive activity on the Sun revolves around the release of magnetic energy into thermal and kinetic energies that propel magnetic structures, called magnetic flux ropes (MFRs), out into the heliosphere. The topology of the coronal magnetic fields is key because it can either enable the release of magnetic energy via reconnection (\citealt{Priest_Demoulin_1995}, see review by \citealt{Longcope_2005}) or provide an escape route to a CME (e.g. \citealt{Antiochos_etal_1999}, \citealt{Aulanier_etal_2000}, \citealt{Lugaz_etal_2011}, \citealt{Sun_etal_2012b}, \citealt{Sun_etal_2013},  \citealt{Jiang_etal_2013}). 

In a 2D magnetized conducting medium, a special location may exist in it, where the magnetic field magnitude goes to zero, i.e. $|\textbf{B}|=0$. Null points are common features of magnetic fields originating from multiple sources, e.g. two adjacent bipoles on the sun. In such a simple 2D quadrupolar configuration the magnetic field connectivity can be organized into four domains, separated by two imaginary intersecting lines, the \emph{separatrices}. Right at the intersection of the separatrices, an X-type null point exists giving rise to the so-called ``X-type'' magnetic configuration. Such topological features tend to be locally unstable when the magnetic sources (e.g. flux concentrations such as sunspots) are free to evolve \citep{Dungey_1953}. This process, known as \emph{X-type point collapse}, leads from an initial equilibrium state to a current sheet $-$ a necessary ingredient for the reconnection of magnetic field lines to occur.

In 3D, a quadrupolar configuration may or may not yield these unstable topological features - the 3D null points. Indeed, 3D null points are unable to explain the variety of observed flaring configurations in Active Regions (AR) on the sun. \citet{Priest_Demoulin_1995} explored a way for magnetic reconnection to occur in 3D in the absence of 3D null points, in electric current surfaces associated with topological features known as the \emph{Quasi-separatrix Layers} (QSLs; see review by \citealt{Longcope_2005}). These finite-thickness current surfaces (hence layers) form when there is an abrupt change in the magnetic field line linkage. This is a typical case in multiple-source AR configurations. The more abrupt the change in the field line linkage, the more intense and thin these current layers are. When two QSLs intersect, they form another topological feature known as a \emph{Hyperbolic Flux Tube} (HFT) \citep{Titov_etal_2002, Titov_etal_2003}. In the limit where QSLs reach infinitesimal thickness, an HFT becomes a line, the \emph{separator}. The separator is a \emph{locus} of X-points in 3D and harbors the strongest topologically-induced electric currents (over which the mapping of magnetic field lines is discontinuous). For completeness, we should also mention here the topological case of a 3D null-point in the corona with a single QSL dome. This dome-shaped QSL (also known as ``the fan'') is manifested by field lines deflecting around (or ``fanning out'' from) the 3D null-point and each of them ending at different points at the base of the dome (at the surface). In this case where there is only one QSL dome there is no HFT/separator, but a special field line passing right through the null point called the ``spine''. 

Since the aforementioned topological features emanate from the instantaneous configuration of magnetic sources in the photosphere, their lifetimes are dependent on the persistence of the relative position of the magnetic sources on the surface. These features may last for days, gradually evolving in timescales similar to the timescales of photospheric motions of the magnetic polarities.

Most of the literature on the initiation of CMEs has focused on the role of topology in \emph{facilitating} solar eruptions. \citet{Aulanier_etal_2000} analyzed an eruptive flare originating from a $\delta$-spot quadrupolar AR which formed after the emergence of a parasitic bipole. Magnetic field extrapolations suggested that the topology was of the ``fan'' and ``spine'' type and that break-out reconnection \citep{Antiochos_etal_1999} at the null-point was the trigger for the destabilization of a filament and its eruption. \citet{Sun_etal_2012b} has reported observations of a non-radial (highly inclined initial trajectory) eruption giving its place to a stable, jet-like inverted-``Y''-shaped structure in its wake. A non-Linear Force-Free Field (in short, NLFFF) extrapolation suggested the existence of a coronal 3D null-point with the wide part of the inverted-Y-shape structure likely due to the lines of the fan separatrix surface. They concluded that this special field geometry guided the non-radial eruption during its observed initial stage. Recently, \citet{Sun_etal_2013} reported another eruption associated with the presence of the same topological features. They associated the triggering of the eruption to break-out reconnection at the null-point. This scenario has been investigated numerically by \citet{Lugaz_etal_2011} (with an out-of-equilibrium initial flux-rope embedded in a ``fan''-``spine'' topology) and \citet{Jiang_etal_2013} (starting with a NLFFF equilibrium). In both cases the null-point reconnection intensified and the fan dome opened up. While the MFR will most likely begin reconnecting with the overlying field, a current sheet forms behind it as it rises giving birth to standard ``flare reconnection''. These works, and much additional literature, suggest that the magnetic field topology over an unstable region plays a facilitating role to the eruption and hence much attention has been devoted in locating such topologies. On the other hand, the question on whether (and how) topology could \emph{prevent an eruption} has not received much attention.

In this work we present evidence for the ``\emph{preventative}'' role  magnetic topology can play for an erupting MFR and explain how it can `kill' the eruption at its initial stage by destroying the rising magnetic structure. The event was captured by a unique and comprehensive set of instruments during the observing campaign in support of the launch of \emph{VAULT2.0} (\citealt{Vourlidas_etal_2016}) on September 30, 2014. 

The paper is organized as follows. In Section~2 we present the observations around the time of the event followed by a detailed observational analysis focused on the dynamic nature of the ``cusps'' in Section~3. Next, we expand on the extrapolation method we used to obtain a model for our target active region's magnetic field and its evolution in Section~4 and then we undertake a topological analysis in Section~5. We discuss our results in Section~6 and summarize and conclude in Section~7.

\section{Campaign Observations}
\textit{VAULT2.0}'s observing window lasted for five minutes (18:09 - 18:13 UT). However, the campaign observations were initiated before and ended much later than \emph{VAULT2.0}'s flight. For this work we used continuous observations from the \textit{Atmospheric Imaging Assembly} (\textit{AIA}; \citealt{Lemen_etal_2011}) onboard the \textit{Solar Dynamics Observatory} (\textit{SDO}; \citealt{Pesnell_etal_2012}) covering the entire solar disk at 0$\farcs$6 pix$^{-1}$ every 12\,s in several coronal passbands (sensitive to different plasma temperatures) $-$ here we used image time series in 304\,\AA\ (50,000K),  94\,\AA\ (6.4 MK) and 131\,\AA\ ($\sim$10 MK). For the photospheric magnetic field distribution we used magnetograms from the \textit{Helioseismic and Magnetic Imager} (\textit{HMI}; \citealt{Scherrer_etal_2012}) also onboard \textit{SDO}. The \textit{SDO} spacecraft orbits the Earth at an inclined geosynchronous orbit. For assessing whether there was a CME associated with our event, we used data from the \textit{Large Angle Spectrometric Coronagraph} (\textit{LASCO}; \citealt{Brueckner_etal_1995}) aboard the \textit{Solar and Heliospheric Observatory} (\textit{SOHO}; \citealt{Domingo_etal_1995}), which observes from the Lagrangian L1 point between the Sun-Earth line. The \textit{Interface Region Imaging Spectrograph} (\textit{IRIS}; \citealt{dePontieu_etal_2014}) provided slit-jaw image time-series in \ion{Si}{4} 1400\,\AA\ with 0$\farcs$16 pix$^{-1}$ every 13\,s and with a field of view (FOV) of 120$\arcsec\times$120$\arcsec$. In addition, the \textit{Solar Optical Telescope} (\textit{SOT}; \citealt{Tsuneta_etal_2008}) onboard the \textit{Hinode\/} spacecraft provided rasters in the \ion{Ca}{2} H 3968\,\AA\ (3\,s time cadence at 0$\farcs$054 pix$^{-1}$ and a FOV of 55$\arcsec\times$55$\arcsec$). The coordinated observation of the target with IRIS/SJI and SOT is a challenging task (in particular, maximizing the overlap between the FOVs of these instruments). In our observational campaign, SOT has 71\% overlap with IRIS/SJI with a common FOV of 44$\arcsec\times$49$\arcsec$.

\begin{figure*}
\epsscale{.9}
\plotone{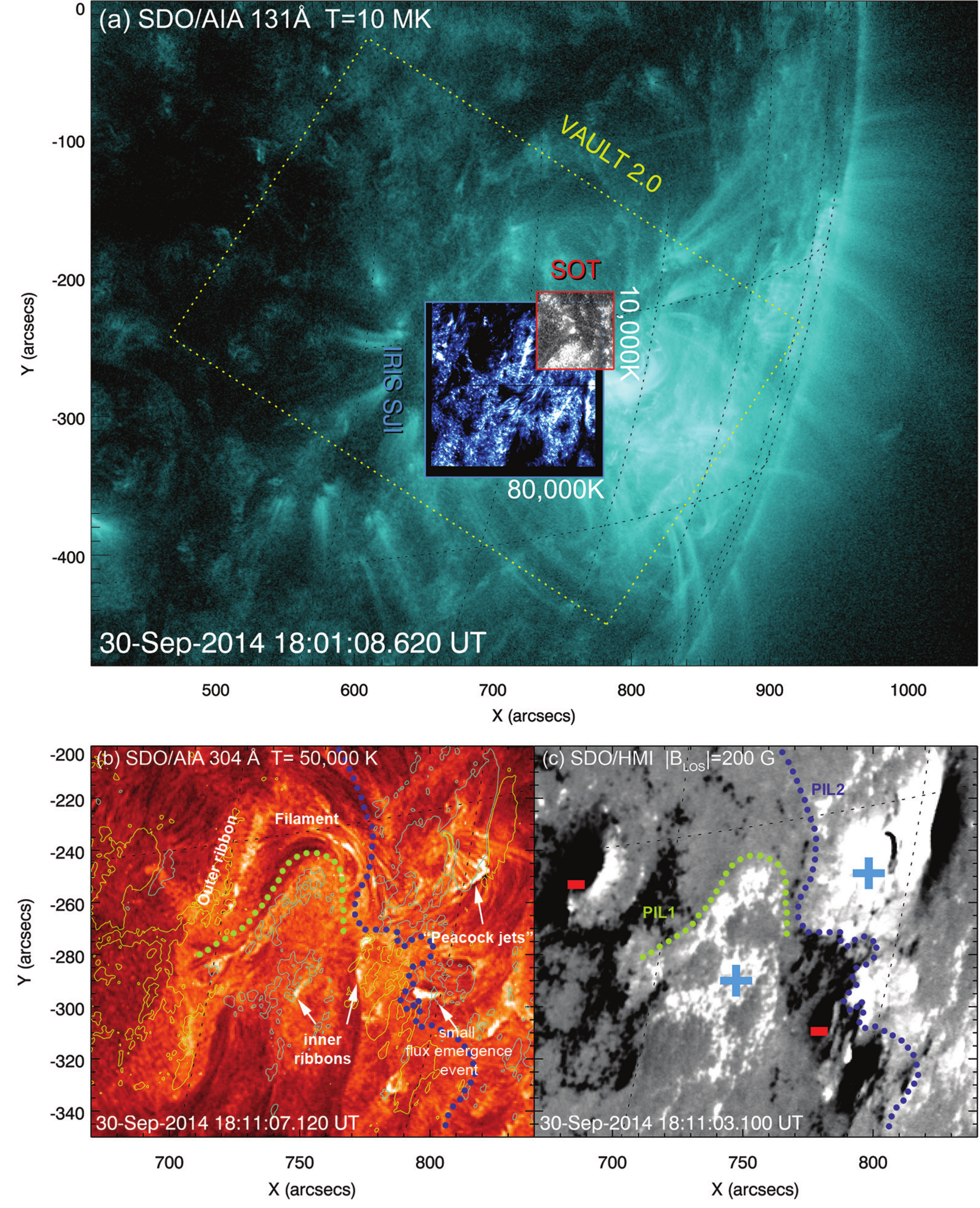}
\caption{(a)  The FOV of each instrument (IRIS Slit-Jaw Imager and SOT) superimposed on a 131\,\AA\ snapshot image from SDO/AIA. (b) Image from SDO/AIA at 304\,\AA\ showing the filament and other annotated features in the core of target AR12172. Contours of photospheric $|\mathrm{B}_\mathrm{LOS}|$=200\,G (positive/light blue; negative/yellow) are overplotted on the 304\,\AA\ image. (c) $|\mathrm{B}_\mathrm{LOS}|$ map from SDO/HMI of the same area of panel (b) showing the general quadrupolar configuration of AR12172. The PILs are traced with colored dots in both panels.}
 \label{IRIS_SOT_AIA_filament}
\end{figure*}

An eruption began over the northern bend of AR12172's inverse ``U''-shape filament at 17:50 UT and was observed by IRIS, SOT and AIA (Figure \ref{IRIS_SOT_AIA_filament} (a-b)). The heliographic position of the target AR on the solar disk was S12$\degr$W54$\degr$. The event exhibited all the typical signatures of an eruption, such as extended ribbons, rising loops, brightenings, coronal rain and several million degree heating. However, there was no CME or any significant radial outflow high in the corona (no such signatures in LASCO Coronagraphs). Hence, it was a failed eruption.

Thanks to the on-disk location of AR12172, observations of the photospheric magnetic field were obtained by the HMI instrument. As seen in Figure \ref{IRIS_SOT_AIA_filament} (c) the photospheric configuration for this AR is quadrupolar, with two Polarity Inversion Lines (PILs). This quadrupolar AR is the result of colision between a strong pre-existing bipole (with PIL1) and a secondary bipole (primarily responsible for causing the East-West portion of PIL2). This colision occured several days before the failed eruption. Furthermore, there is a small tertiary emergence event within the secondary bipole in PIL2 (Figure \ref{IRIS_SOT_AIA_filament} (b)) lasting throughout our observations. 

The inverse U-shaped filament is in fact sitting above the PIL1 but cool material is also seen above PIL2. In actuality, the southwestern part of the inverse-U shape filament seems to be inactive at all times, possibly due to the strong magnetic fields about that part of the PIL1. Images taken with SDO/AIA in 304\,\AA\ and IRIS SJI in 1400\,\AA\ suggest that the southwestern part of the filament is not continuous with the rest of the filament. Thus, the relevant length of the filament reduces its inverse U-shape into a forward ``S''-shape (as delineated by PIL1 in Figure~\ref{IRIS_SOT_AIA_filament} (b-c)). Over time, brightenings and flows suggest that this shape extends towards the west, over PIL2 (more below).

\begin{figure*}
\epsscale{.90}
\plotone{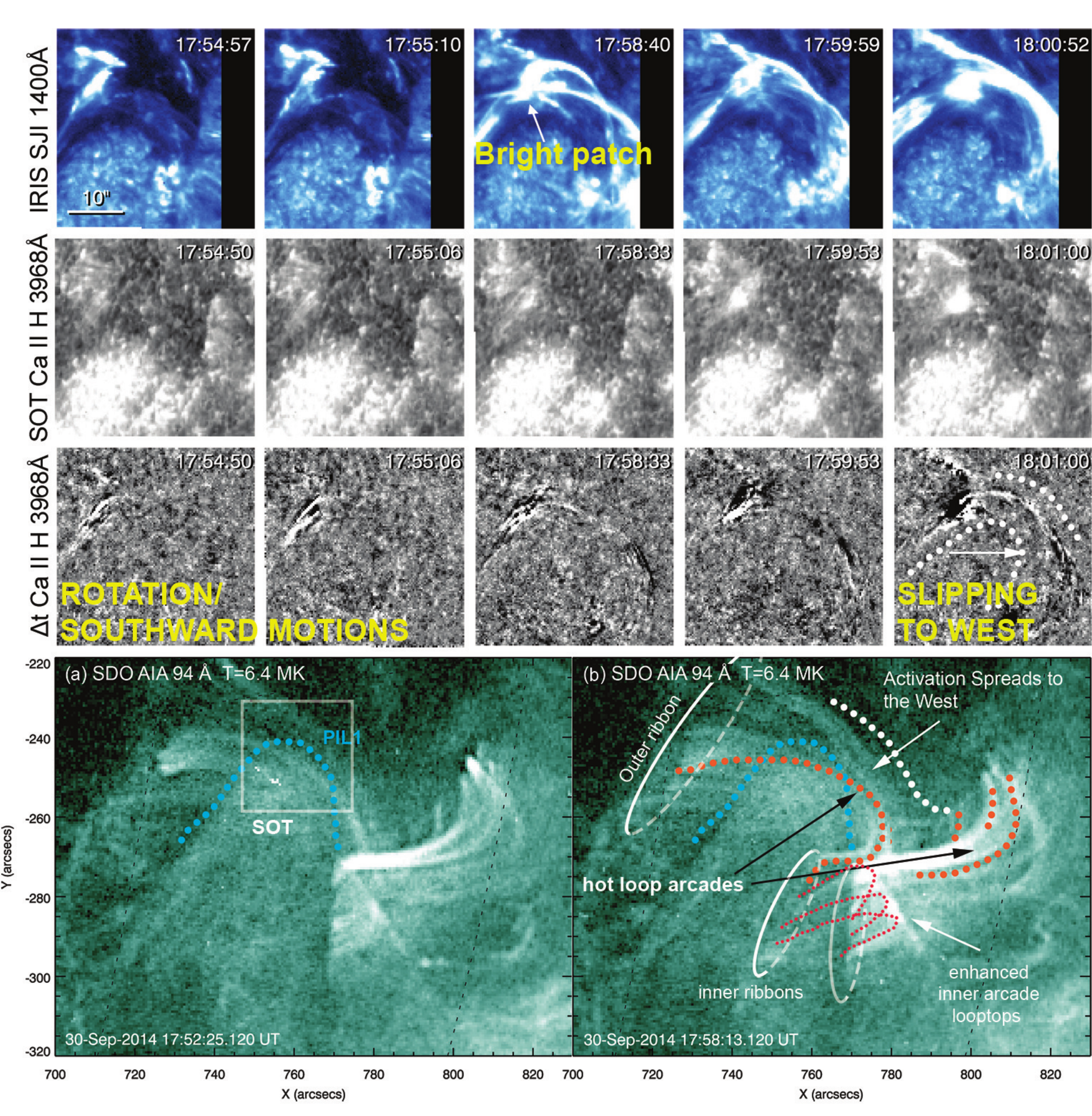}
\caption{Snapshots during the time of the filament activation. The activation appears to begin with rotation/unwinding of bright structures above the N part of the filament. The brightenings spread above and along the length of the filament (shown here over a period of 6 min) with apparent rising of these bright structures (a running difference movie of the SOT observations is available in the online version of the journal.)  Bottom panels: (a) The response in the corona is instantaneous, with hot loops brightening in 94\,\AA\ passband of AIA (6.4\,MK). (b) 6 mins later, as the connectivities seen in SOT and IRIS extend westward, a hot (6.4\,MK) hook-shaped post-flare loop arcade is seen in the East above the filament (94\,\AA ; orange dots). The propagation of the activation towards the West is also reflected in 94\,\AA\ (long connectivities above PIL2; below white dots) in addition to the bright hot arcade in the West (orange dots).}
\label{IRIS_SOT_AIA_evolution}
\end{figure*}

The SOT observations at the \ion{Ca}{2} H 3968\,\AA\ line (small FOV raster scans at an ultra fast 3\,s cadence) are centered on the north bend of PIL1. In these observations the filament is invisible (although it is seen in IRIS 1400\,\AA\ and AIA passbands). During the time of the first filament brightenings ($\sim$ 17:52 UT) a location above the north bend of PIL1 shows ``finger''-like structures to slowly brighten while they exhibit rotational/unwinding behavior. A few snapshots showing these structures are presented in the SOT panels of Figure \ref{IRIS_SOT_AIA_evolution}. Despite the strong background photospheric emission, these dynamic structures stand out in the corresponding running-difference image series ($\Delta t$ \ion{Ca}{2} panels; Figure \ref{IRIS_SOT_AIA_evolution} and ``movie1.mp4"). These apparent rotational motions occur above the PIL1 where the filament resides and within a few minutes extend well along the entire north segment of the PIL1 and neighboring PIL2. In addition, these bright structures seem to be moving upwards as suggested by their relative northwest shift with respect to the dark filament which remains still in the background. \ion{Ca}{2} brightenings are usually associated with transient heating of chromospheric material or microflare reconnection in the corona \citep{Shimizu_2011}. 

This extension of activity along the PILs is accompanied by similar motions and brightenings in transition region temperatures ($\sim$80,000\,K seen in IRIS \ion{Si}{4} 1400\,\AA) suggesting the heating of structures right above the dark filament. In addition, the bright SOT ``finger-like'' structures seem to migrate westwards towards the neighboring PIL2 and seem to be related to a bright patch in \ion{Si}{4} 1400\,\AA\ that behaves similarly. Interesting connectivity seems to link the bright patch to other locations along the filament (IRIS SJI panels 17:58:40 - 18:00:52 in Figure \ref{IRIS_SOT_AIA_evolution}). 

In tandem with the activity in the chromosphere and the transition region (TR), there was a coronal response with multi million-Kelvin plasma emission above the filament (Figure \ref{IRIS_SOT_AIA_evolution} (a)). A movie covering the evolution in the corona and TR is available in the online version of the journal (``movie2.mp4''). As seen in 94\,\AA\ (6.4 MK) a narrow loop arcade over the north bend of the filament from NE to SW brightens in sync with the chromospheric/TR brightenings (i.e. SOT ``fingers'' and their IRIS counterparts). In fact, the narrow arcade's footpoints correspond to bright ribbon-like pairs in 1400\,\AA\ from IRIS (17:52:06 1400\,\AA\ panel, Figure~\ref{Cusp_panels_and_slitstackplot}; also in 1600\,\AA\ and 304\,\AA\ from SDO/AIA but not shown here). At the time of the appearance of the bright patch in 1400\,\AA\ (17:58) the narrow loop arcade has extended towards the south in the fashion of post-flare loop arcades (hot hook-shaped East loop arcade and bright West arcade; Figure~\ref{IRIS_SOT_AIA_evolution} (b)), accompanied by extended ribbon brightenings at the footpoints of those loops (e.g. 18:00:12 1400\,\AA\ panel Figure~\ref{Cusp_panels_and_slitstackplot}).

After its launch at 18:09 UT, \textit{VAULT2.0} observed the declining phase of the eruption and captured cooling downflows and bright ribbons (also seen in 304\,\AA\ and 1400\,\AA). This is the first time, to our knowledge, that such a wide range of instruments captured the initiation of an eruption.

\section{Dynamics - Rapid Loop Motions in the 10\,MK Corona}
\begin{figure*}
\epsscale{1.0}
\plotone{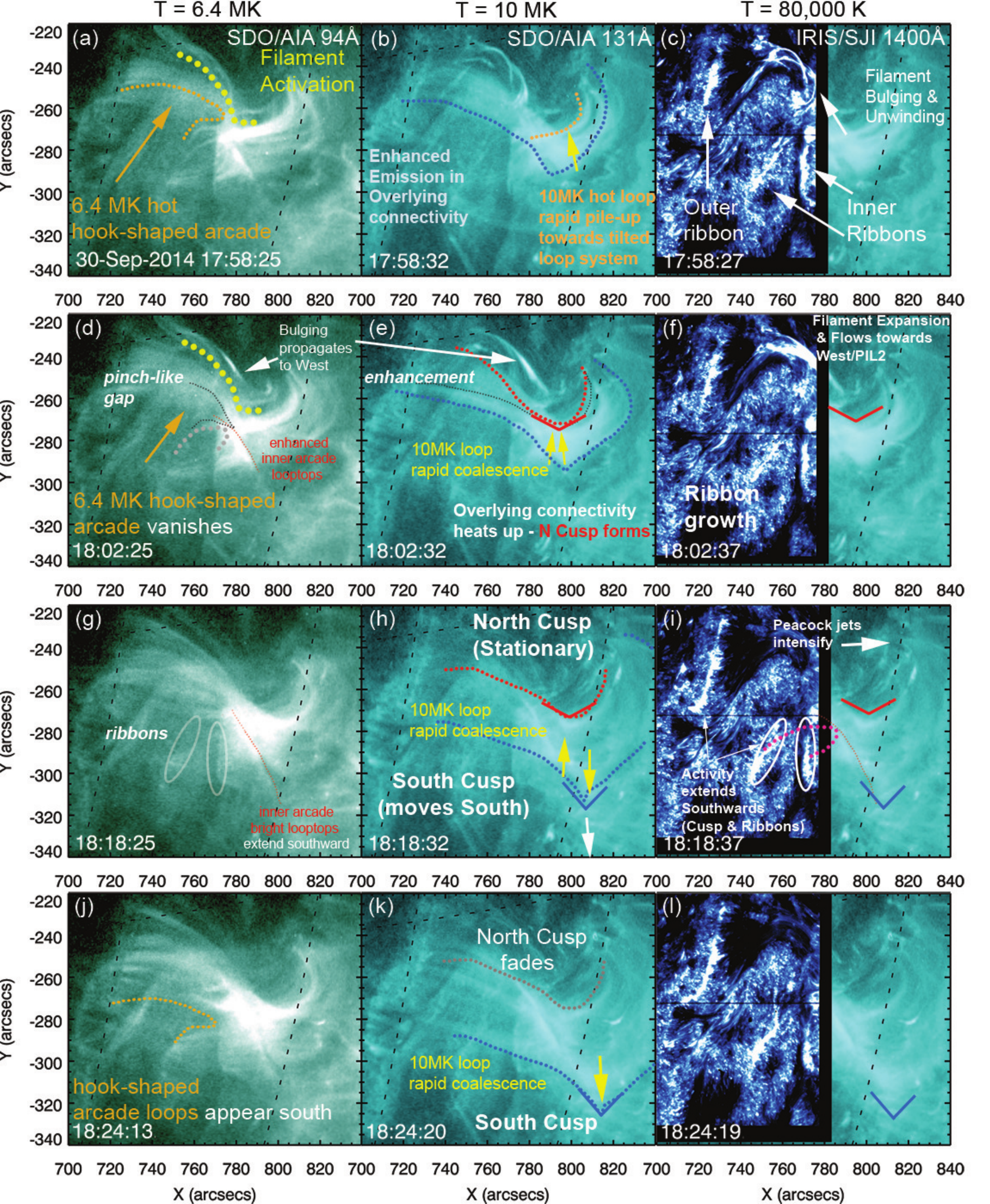}
\caption{Multipanel figure organized in wavelengths (columns of 94\,\AA, 131\,\AA\ and 1400\,\AA) at four different times (rows) summarizing the observations. The small FOV of 1400\,\AA\ is complemented by the respective 131\,\AA\ images. See text for discussion. A movie of this figure is available in the online version of the journal (``movie2.mp4'').}
\label{Arcade_evolution_cusps}
\end{figure*}

As we already mentioned, the corona responded instantly to the activation of the north bend of the filament (filament bulging). A system of ribbons brightens in sync with the activations, with the ribbon showing the fastest growth towards the south (Figure~\ref{Arcade_evolution_cusps} (c)). The ribbon brightenings correspond to the footpoints of the hot loop arcades, namely the 6.4\,MK hot hook-shaped arcade in the East and a bright arcade in the West (Figure~\ref{Arcade_evolution_cusps} (a)). The East hook-shaped arcade shows up almost instantly indicating of very rapid loop motions suggesting rapid pile-up of newly formed loops ($\approx$ 400\,km\,s$^{-1}$).

Just two minutes after the appearance of the bright East and West arcades, the East hook-shaped arcade disappears from the 94\,\AA\ passband leaving a ``pinch''-like gap behind it (Figure~\ref{Arcade_evolution_cusps} compare (a) and (d)). In addition, the East hook-shaped arcade doesn't appear in any other coronal passband. This rapid disappearance is not accompanied by evidence of cooling or heating of these structures given the short period of time. Also, we don't expect to have significant heating (i.e. T$>$10\,MK) given the small magnitude of the X-ray flux (C2.0 level). Note that at that time (18:02), the bulging of the filament has propagated to the West, above PIL2. We observe brightening of structures above and across the PIL of the filament in 94\,\AA\ and 131\,\AA\ (Figure~\ref{Arcade_evolution_cusps} (d) and (e) respectively) together with flows along those bright structures (panel (f)). At the time of the disappearance of the East hook-shaped loops, the East-West overlying connectivity (that seems to link the outer ribbon with the leading positive sunspot, therefore, a large-scale connectivity) intensifies, suggesting very high temperatures (131\,\AA\ around 10\,MK). The dissapearance of a hot system and the rapid appearance of another hot system suggests the possibility that these loop systems are interacting through reconnection.

Between 18:02 to 18:18, the ribbons and the 10\,MK large-scale East-West loop arcade extend southward. Overall, the East-West axis of this large-scale arcade appears to follow the shape of the filament channel (Figure~\ref{Arcade_evolution_cusps} (a) for the filament; also compare the red and blue dotted envelopes in the 10\,MK panels (e) and (h)). The middle part of this East-West curvature can be described as ``V''-shaped (convex part pointing Southward), with its North envelope/boundary remaining largely in place and the South envelope growing southward following the ribbon growth. 

Immediately after the disappearance of the 6.4\,MK hook-shaped arcade at 18:02 UT, a ``V-cusp'' shaped structure appears to match the convex-up part of the ``V''-shaped North envelope of the 131\,\AA\ East-West loop system (Figure~\ref{Arcade_evolution_cusps} (e) thick red V-shape). Interesting activity accompanies this cusp: bright loop structures appear suddenly near the convex side of the cusp (i.e. just South of it) coalescing with the cusp. This rapid hot loop coalescence enhances the emission in the overlying arcade (Figure~\ref{Arcade_evolution_cusps} (e); dark dotted line) and also at the ribbons lower in the atmosphere (Figure~\ref{Arcade_evolution_cusps} (f)). Since this ``V''-cusp (and the rapid loop coalescence) is seen only in 131\,\AA\ it likely corresponds to $\approx$ 10\,MK plasma. This cusp seems to be stationary at all times, suggesting its collocation with the also stationary North envelope of the East-West overlying arcade. We dub this ``V''-cusp, the ``North cusp''. 

At the south envelope, another ``V''-cusp seems to form within minutes (18:04 UT) from the first appearance of the North cusp, with its convex part pointing to the same southward orientation. We dub this the South cusp. The South cusp apparently leads the expansion/growth of the overlying hot EW loop arcade (Figure~\ref{Arcade_evolution_cusps}; compare blue-dotted envelope in panels (e), (h) and (k)), essentially marking its position as a bright front. The rapid coalescence of bright loops towards the cusps continues with loops coalescing to the convex side of the North cusp and loops coalescing towards the concave part of the South cusp (pointed by the yellow arrows in Figure~\ref{Arcade_evolution_cusps} in panels (b), (e), (h) and (k)). Note that the loop coalescence at the South cusp is much faster than the apparent southward migration of the South cusp. As the South cusp migrates south, hook-shaped loop arcades gradually reappear in the East, seemingly anchored to the elongated ribbon and the East of the inner ribbons. At around 18:18 UT the loop coalescence towards the North cusp ceases and the cusp fades. The South cusp continues to be active up until 18:40 UT, where it fades as well. The observations we described up to here are organized in three stages (I, II and III) in Table~\ref{TABLE}.

\begin{figure*}
\epsscale{0.90}
\plotone{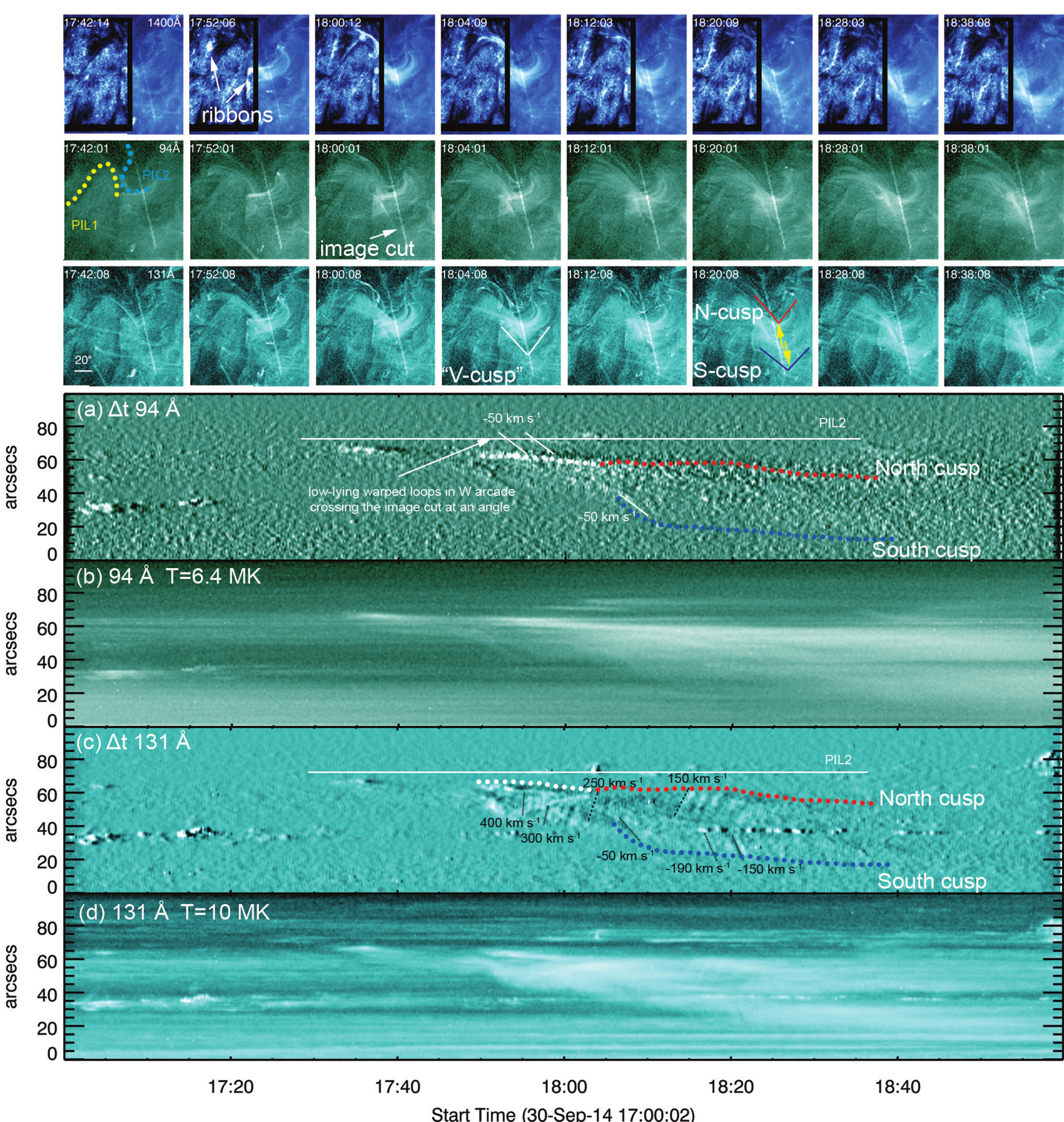}
\caption{Upper panels: Top row: IRIS 1400\,\AA\ with 131\,\AA\ complementing the non-observed area by IRIS. Bright ribbon structures extend southwards during and following the filament activation. Middle row: AIA 94\,\AA\ (T $\approx$ 6.4 MK). Bottom row: AIA 131\,\AA\ (T $\approx$ 10 MK). A narrow loop ``bundle'' above the filament brightens and expands southward. ``V''-like cusps form, both leading a ``post-flare-loop''-like expansion anchored to the ribbons. Hot loops are seen to coalesce rapidly at the cusps (only seen in 131\,\AA\, signifying that they are very hot structures). The position of the image cut used in the time-distance analysis is shown as a transparency on each image. Bottom panels: Space-time plots for 94\,\AA\ and 131\,\AA\ and their time derivatives ($\Delta t$). The fast loop coalescence towards the cusps (red and blue dotted envelopes) is best seen in 131\,\AA\ $\Delta t$ revealing a ``herring bone'' pattern which indicates motions towards N and S direction along the slit. Selected slopes are annotated with apparent speeds in km s$^{-1}$. The South cusp migrates towards the south, while the North cusp remains more stationary. A running difference movie showing the cusps in 131\,\AA\ is available in the online version of the journal (``movie3.mp4'').}
\label{Cusp_panels_and_slitstackplot}
\end{figure*}

By extracting a fixed 100$\arcsec$-long linear strip (or ``cut'') of pixels in each 131\,\AA\ and 94\,\AA\ frame of the image series by SDO/AIA and by stacking each strip along the time-dimension (i.e. space-time plot), we can measure the speed of the coalescing bright loops along the dimension of the strip (see bottom space-time plots in Figure \ref{Cusp_panels_and_slitstackplot}). The linear cut intersects both cusps. 

As we have already mentioned, the North/South cusps and the coalescing loops are seen in 131\,\AA\ suggesting temperatures of 10\,MK in these structures. However, as it can be seen in panels (b) and (d) of Figure~\ref{Arcade_evolution_cusps}, the cusp structures are faint (especially at later times) but also are seen against strong background emission, either due to hot and bright structures at lower heights in 131\,\AA\ but also due to a cool line component (\ion{Fe}{8}; 400,000\,K) near the 131\,\AA\ \ion{Fe}{23} (10\,MK), which also contributes to the passband. The latter is not an issue in 94\,\AA\ images as they predominantly show plasma emission at $\sim$6.4\,MK. However, the 94\,\AA\ image series is only able to capture traces of the bright fronts of the cusps, but not the loop coalescence towards the cusps. To improve the visibility of the imprints the moving features leave in the space-time plots, we also show the time-derivative (Figure \ref{Cusp_panels_and_slitstackplot} (a) and (c)). A two-panel movie with both the direct 131\,\AA\ observations and their running difference is available in the online version of the journal (``movie3.mp4''.)

Starting with the hot loop rapid pile-up in the W arcade at 17:52-17:58 UT (Figure~\ref{Arcade_evolution_cusps} panels and space-time plot and Figure~\ref{Arcade_evolution_cusps} panel (b)) their estimated plane-of-the-sky speeds are $\sim$ 400\,km s$^{-1}$ (positive speeds are Northward along slit, negative southward), which occur at the time when the E hook-shaped loop arcade first appears in 94\,\AA\ (17:53 UT; Stage I in Table~\ref{TABLE}). From the way the loops are bent, it seems that the Northward motions are essentially downward motions, hence the characterization ``loop pile-up''. The speeds drop to 300\,km s$^{-1}$ at 18:00 UT. This Stage I activity ceases with the disappearance of the hook-shaped loops seen in 94\,\AA. 

At 18:02 UT (Stage II in Table~\ref{TABLE}) we have the formation of the North cusp by rapid coalescence of loops with speeds of  $\sim$250\,km s$^{-1}$. Simultanesously to the coalescence of loops, the hook-shaped loop arcade seen in 94\,\AA disappears. Not too late after the formation of the North cusp, the South cusp appears to move southward with -50\,km s$^{-1}$ for about 4 minutes; a speed suggestive of a slow quasi-static evolution. After this initial southward expansion, the South cusp moves southward with $\sim$ -5\,km s$^{-1}$. Contrary to the initial evolution seen in the North cusp, the initial (first 4 minutes) loop coalescence speed in the South cusp is comparable to the latter's proper motion, i.e. between 60-75\,km s$^{-1}$. At 18:18 UT, the loop coalescence speed for the North cusp dropped to 150\,km s$^{-1}$, while the South cusp coalescence speed reaches a maximum of -190\,km s$^{-1}$. 

Later on, at 18:20 UT (well into Stage III, Table~\ref{TABLE}), the North cusp loop coalescence stops, while coalescence continues in the South cusp with -150\,km s$^{-1}$ and with a southward proper motion for the cusp/front at $\sim$ -5\,km s$^{-1}$. The speed of loops of $\sim$ 150\,km s$^{-1}$ is a fraction of the Alfv\'{e}n speed at intermediate heights and magnetic field strength in the corona ($v_A\approx1,000$\,km s$^{-1}$).

\begin{deluxetable}{rccc}
\tablecolumns{4}
\tablewidth{0pc}
\tablecaption{Summary of Observed Failed Eruption.\tablenotemark{*}}

\tablehead{
\colhead{Temperature ($\lambda$)} & \colhead{STAGE I (17:52-18:02)} & \colhead{STAGE II (18:02-18:15)} & \colhead{STAGE III (18:15-18:40)} }
\startdata
10\,MK (131\,\AA) & W hot loop & hot loop rushing & N-cusp rushing ends \\
       & rapid pile-up & to N/S-cusp & S-cusp extends S \\ 
6.4\,MK (94\,\AA)  & E hook-shaped arcade & E hook-shaped arcade & E hook-shaped arcade \\
   & appears  &  vanishes, ``pinch''-like gap & reappears extending S \\ 
80,000\,K (1400\,\AA) & filament bulging at N bend/ & Bulging propagates W/ & Extended outer/\\
   & outer ribbons & extended outer ribbons & inner ribbons\\

\enddata
\tablenotetext{*}{Times are given in Universal Time (UT)}
\label{TABLE}
\end{deluxetable}

The occurrence of two slowly moving convex 10\,MK cusps is an interesting coronal phenomenon which requires further investigation. This ``V''-shape loop structure (which as we said occurs at two distinct locations) cannot be maintained in static conditions due to the restoring effect of the magnetic tension along the loops. However, the magnetic topology may play a role in forging such magnetic structures close to locations of topological features (e.g. QSLs, HFT, etc). 

\begin{figure*}
\epsscale{1.0}
\plotone{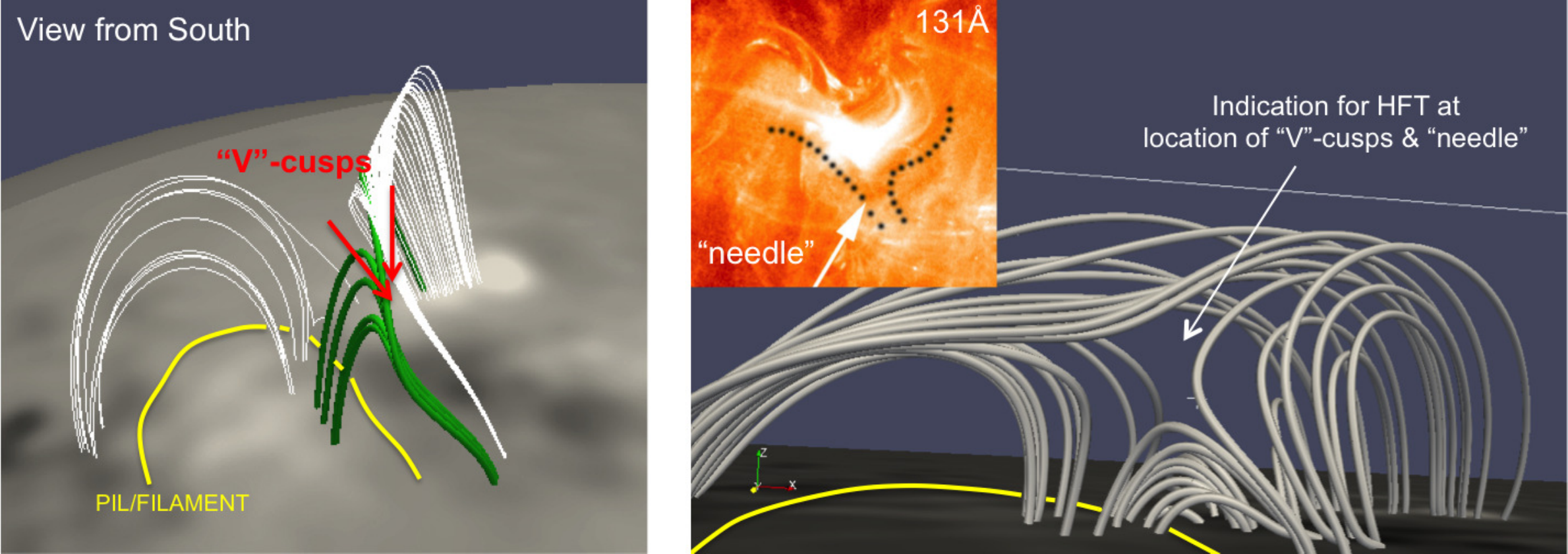}
\caption{Left: Result of the PFSS extrapolation with boundary conditions appropriately updated for the area of AR12172 at 2014-09-30 17:34 UT ($\sim$30 mins before \emph{VAULT2.0} flight). In this view the field is visualized in spherical coordinates. Selected Green lines show lines converging southwards from the East and white lines converging southwards from the west suggesting the existence of a topological feature recovered in our PFSS extrapolation. Lines that seem to be converging from the East and West sides of the AR yield an apparent ``V''-cusp morphology around the location of convergence. Right: Extrapolation viewed in Cartesian coordinates from the south; note the location of the ``X-type'' configuration. Inset image of 131\,\AA\ shows the cusp morphology (dotted lines) and an apparent ``needle''-like structure, similar to a non-radial inverse-``Y''-shaped feature reported by \citet{Sun_etal_2012b}. 
(The view for both figures is from the South towards the North.)}
\label{FIG_PFSS_CUSP_XPOINT}
\end{figure*}

Indeed, the existence of a topological feature at the observed location of the cusps is suggested by a Potential Field Source Surface extrapolation (PFSS; \citealt{Schatten_etal_1969}) around the time of the event (2014-09-30 17:34 UT; Figure \ref{FIG_PFSS_CUSP_XPOINT}). For the boundary conditions of the PFSS extrapolation we updated a synoptic magnetogram map for Carrington rotation 2155 by patching a 10$\degr\times$10$\degr$ magnetogram (appropriately mapped to Carrington heliographic coordinates). The PFSS extrapolation was performed with 330 orders for the spherical harmonics. As viewed from the south (left and right panel of Figure \ref{FIG_PFSS_CUSP_XPOINT}), the structuring of magnetic field lines seems to follow an X-type null-point topology, suggesting the existence of an HFT oriented North to South along the ``X-type'' topology above the filament.

There is another interesting ``needle''-like feature associated with the 131\,\AA\ cusps (inset image of Figure \ref{FIG_PFSS_CUSP_XPOINT}). This ``needle''-like structure is very similar to a non-radial inverse-``Y''-shaped feature reported in the work of \citet{Sun_etal_2012b} right after a filament eruption in a fan-spine topology. Despite this morphological (or topological?) similarity, there was no CME associated with our event. The ``needle'' structure is seen only in 131\,\AA\ (10\,MK plasma emission) suggesting that is associated with excess plasma heating along the separator/HFT (corresponding to the extrapolated field lines in Figure \ref{FIG_PFSS_CUSP_XPOINT}).

Indications for the topological features we seek were found reasonably close to their expected positions, given the assumptions of the PFSS model. However, the PFSS model is by construction not able to model current-carrying structures in the corona (e.g. a filament), not to mention that it is a static model and it does not capture the dynamic evolution. The activity and the dynamics we observed in this AR involves a filament, and thus it is necessary to consider its presense when modeling the coronal magnetic field. Non-linear Force-Free Field (NLFFF) extrapolations have been successful in reproducing filaments and strong indications for MFR-like structures (e.g. \citealt{Chintzoglou_etal_2015}; also see review by \citealt{Wiegelmann_Sakurai_2012}). This class of models implements the full magnetic vector at the photosphere for the lower boundary condition. Vector magnetograms, while nowadays routinely obtained by the HMI instrument in high spatiotemporal resolution, suffer from instrumental effects (namely high noise levels for the transverse component of the magnetic field increasing with the distance from the center of the solar disk). Indeed, due to the position of AR12172, which is off the center of the disk (S12$\degr$W54$\degr$), the noise levels in the vector magnetograms are high and the filament cannot be realistically reproduced from the vector boundary conditions by a NLFFF extrapolation method. Representing the filament in the model is necessary in order to investigate its interaction with the magnetic topology of its host AR. An NLFFF model that can account for the aforementioned limitations is the Magnetic Flux Rope-insertion method (\citealt{vanBallegooijen_2004}). We discuss this model in the following section.

\section{NLFFF Modelling of the event with the MFR-insertion method}

The purpose of the NLFFF modeling is to determine whether the observed structures seen in the chromosphere (filament) and the corona (brightenings \& interaction of an MFR with the overlying field topology) can be explained in terms of stable quasi-static models (as suggested by the relatively slow evolution of the event). NLFFF extrapolations (e.g. \citealt{Wiegelmann_2004}) and potential (\citealt{Schmidt_1964}) extrapolations, while capable of capturing the topological structures, are merely ``static'' models, incapable to illustrate any evolution of the system.

In order to capture the dynamic interaction in the system we used the flux-rope insertion method (\citealt{vanBallegooijen_2004, Savcheva_vanBallego_2009, Su_etal_2009, Savcheva_etal_2012, Bobra_etal_2008}). The 3D magnetic field of the solar corona is modeled using line-of-sight (LoS) magnetograms. An MFR is inserted in a high-resolution potential field extrapolation along the path of the observed filament in AIA 304\,\AA. The configuration is relaxed to a force-free state by magnetofrictional relaxation (\citealt{Yang_etal_1986, vanBallegooijen00}). Magnetofriction (MF) assumes that the Lorentz force in the corona acts against an ad-hoc frictional force. 
MF consists in evolving the coronal field via the ideal induction equation, expressed in terms of the vector potential:

\begin{equation}
\dAdt = \Bv\times\BB+\eta\Bj+\frac{\BB}{B^2}\nabla\cdot(\eta_4B^2\nabla\alpha), 
\end{equation}
where $\mathbf{A}$ is the vector potential, $\BB=\nabla\times\BA$, $\Bj=\nabla\times\BB$ is the current density, and $\Bv$ is the magnetofrictional velocity, $\Bv=\frac{1}{\nu}\frac{\Bj\times\BB}{B^2}$, with $\nu$ the coefficient of friction. The coefficient $\eta$ is the ordinary diffusion. $\eta_4$ is the 4-th order diffusion, called hyperdiffusion \citep{vanBallegooijen00}, which acts to smooth gradients in the torsion (force-free) parameter, 

\begin{equation}
\alpha=\frac{\Bj\cdot\BB}{B^2}, 
\end{equation}
which for a NLFFF is constant along a given field line but allowed to vary from field line to field line. For a NLFFF, the induction equation is iterated until the MF velocity vanishes as per the assumption that the configuration is in equilibrium for a stable NLFFF model. For an unstable model there is a residual Lorentz force and the velocity does not reach zero, meaning that the MFR continues to evolve with each subsequent iteration of the MF equation. The uniqueness of the method lies in the ability to produce viable unstable models. 

A grid of models with different combinations of poloidal and axial flux in the subject MFR was constructed. The next step would be to match field lines from the models to observed coronal loops in X-rays and EUV, thus finding the best candidate model. However, in this case we model an unstable configuration, and have no suitable AIA or XRT loops to use so we use the elongated brightenings that appear as the filament material is activated.   

\citet{Bobra_etal_2008} and \citet{Su_etal_2011}  found that the MF relaxation has two possible outcomes. Either the MFR settles into a force-free state, or the field expands indefinitely  never reaching a force-free state. This loss of equilibrium occurs when the axial flux is larger than a certain value (the ``stability limit''). This means that the MFR-insertion method is the appropriate model to use in order to capture the dynamics of a quasi-static evolution inferred from the observations. \cite{Savcheva_etal_2015a, Savcheva_etal_2015b} showed the usefulness of using unstable models to represent erupting configurations. 

For our investigations we produced MFR models of the following characteristics: (a) a stable MFR, with axial flux of $1\times10^{21}$\,Mx and poloidal flux $5\times10^{9}$\,Mx\,cm$^{-1}$, and (b) an unstable MFR, with axial flux $2\times10^{21}$\,Mx and poloidal flux $5\times10^{10}$\,Mx\,cm$^{-1}$. The stable model took 40,000 iterations to converge, while the unstable model never achieved convergence. Instead, the unstable model evolves via MF where it inflates and interacts with the HFT towards the South (see next section). For our investigation, we stored several selected snapshots during the MF relaxation of the stable (snapshot of iteration 40,000) and unstable model (snaphsots of iteration 10,000, 20,000, 30,000, 45,000).

\section{Topological Analysis}

One way to extract physically relevant information from complex magnetic field models is to analyze their magnetic topology. This is done by grouping field lines into separate bundles which connect disparate regions on the solar surface. These domains are bound by separatrices or QSLs. \citet{Savcheva_etal_2012b} presented the first topological analysis of a NLFFF magnetic field constrained by observations, demonstrating that topological analysis is extremely useful for pinpointing the probable sites of reconnection. Thus, after obtaining the NLFFF for this AR, we can compute the squashing factor $Q$ that defines the strength of QSLs \citep{Titov_etal_2002,titov07}. The $Q$ value is computed by tracing two closely-spaced field lines at one end, and measuring the distance between their conjugate footpoints, as defined by 

\begin{equation}
	Q=\left(\sum_{i,j=1}^2 \left(\frac{\partial X_i}{\partial x_j}\right)^2\right)/|B_{z,0}/B_{z,1}|, 
\end{equation}
where $X_i$ is the coordinate of the conjugate footpoint, and $B_{z,0}$ and $B_{z,1}$ are the vertical field strength at the two footpoints of a field line. This implies the tracing of massive numbers of field lines on a mesh, the resolution of which can be more than an order of magnitude higher than the original magnetic field data. The QSL computations are therefore relying on intensive field line tracing computation. 

To address this computationally challenging problem, an efficient line-tracing code is implemented. It is capable of calculating the $Q$ factor for about a million field lines per minute within a typical AR making use of graphics processing units (GPU). This allows us to explore the topology of regions of interest in two as well as three dimensions with the necessary accuracy (Q reaching values as high as $\sim$10$^{10}$). The method we employ is based on \citet{titov07} and \citet{pariat12}. The method is described in detail in \cite{Tassev_and_Savcheva_2017}.

The calculation of Q was performed for the entire volume of the selected ``snapshot'' cubes for both the stable and the unstable models. In our analysis we visualize these Q-factor 3D cubes in an ``optically thin'' fashion to allow for simultaneous plotting of field lines from the corresponding MF magnetic field cubes. In all of the cases, the overall 3D topology, arising from the (largely) quadrupolar photospheric magnetic field distribution, comprises two 3D QSL ``dome'' structures (QSL1 in the East, QSL2 in the West of the AR) in such close proximity to each other that intersect in the corona. Their intersection gives birth to an HFT, essentially a volumetric structure in the shape of an arc with a cross section resembling the letter ``X'', where each slant of X (``\textbackslash'' and ``/'') represents the cross section of each QSL around the location of their intersection. Due to the relative position of the QSLs, their intersection results to an HFT essentially running along the N-S direction. The height of this HFT structure runs from $\approx$10\,Mm at its North end (i.e. at W end of PIL1) to 45\,Mm (middle/apex) and then dissolves back down to 20\,Mm at its South end.  

In addition to QSL1, QSL2 and HFT structures, there is a QSL structure associated with the inserted MFR, which modulates and intersects/merges with the two aforementioned intersecting QSL domes. This multi-QSL topology makes our analysis a challenging task. The MFR has been inserted along the North part of PIL1 and PIL2 as delineated by the filament structure at the north of the AR core. This PIL is formed by the opposite quiet sun flux elements and the four dominant polarities of the AR. The MFR is initially contained lower than the maximum height of the two QSLs. In the unstable models, the MFR inflates and so is the MFR-associated QSL. Furthermore, there are two QSLs around the MFR, an interior one which exists due to the MFR and an exterior one which exists due to the incompatibility of the expanding arcade around the MFR with respect to the ambient field. The latter MFR-binding QSL expands as MF relaxation proceeds. 

\section{Discussion}

The evolution in the low corona exhibits all the characteristics of a successful eruption, yet there is no CME. The AR contains two PILs that are hosts to a filament (seen as an absorption feature in most wavelengths), which at some point during our observing run brightens along the North bend of PIL1. Then, these bright structures seem to rise while the cool material remains largely static low in the atmosphere. In the Ca II rasters, fine, thread-like structures seem to peel-off above the filament, initially at the north bend and progressively along its length towards the West. Some bright patches show up along these thread-like structures. In coronal/TR wavelengths, the North bend seems to bulge and this bulging continues to PIL2. This succession of events, together with the hot ($\geq 10$\,MK) moving structures (cusps) at larger heights, suggest that magnetic reconnection was driven higher in the corona by these ascending magnetic structures associated with the activity observed above the filament channel. 

Sheared magnetic arcade-like structures may expand upwards with increased shearing of their photospheric footpoints \citep{Mikic_Linker_1994}. However, there is no indication for significant photospheric shearing motions during the short timescale of our event ($\approx$ 1\,h, in total). So, it is tempting to consider the possibility of an ideal MHD instability of an MFR-like filament channel as the driver of the observed activity. In this view, the cool filament material is suspended on the convex-up field lines of a twisted MFR structure. The MFR structure is primarily axial as the filament follows the PIL2 very closely. This is suggestive of a weakly twisted MFR (justifying our choice of a weakly twisted unstable MFR model). A low amount of twist in the MFR rules out the possibility for the Kink instability \citep{Torok_Kliem_2005} and favors the Torus instability, which has a lower twist threshold \citep{Kliem_Torok_2006}.

Evidence for the existence of an HFT in the observations comes from the brightening of loop tops anchored in the two inner ribbons (Figure~\ref{IRIS_SOT_AIA_evolution} (b)). When the inner ribbons are bright, so are the loops anchored in them, presumably delineating the inner arcade that occupies the volume under the two intersecting QSLs. Indeed, the dynamics we observe, namely the moving cusps, occur just above those looptops. In fact, when the South cusp migrates southward, the rapid loop coalescence and the proper motion of that South cusp follows the apparent height of these loop tops. These loop tops are seen best in 94\,\AA\ since they are not obstructed by the overlying hot South cusp (also see ``movie2.mp4'').

This in fact explains the time evolution of the cusps: the first cusp shows up when the MFR begins to inflate (lower height) and the second cusp follows when parts of the MFR expand above the HFT at higher heights. The split of the cusps (Figure~\ref{Cusp_panels_and_slitstackplot}) is not a real split, but an apparent one due to LOS projection of these (quasi-)static (i.e. HFT) and dynamic (i.e. MFR) systems.

\begin{figure*}
\epsscale{1.0}
\plotone{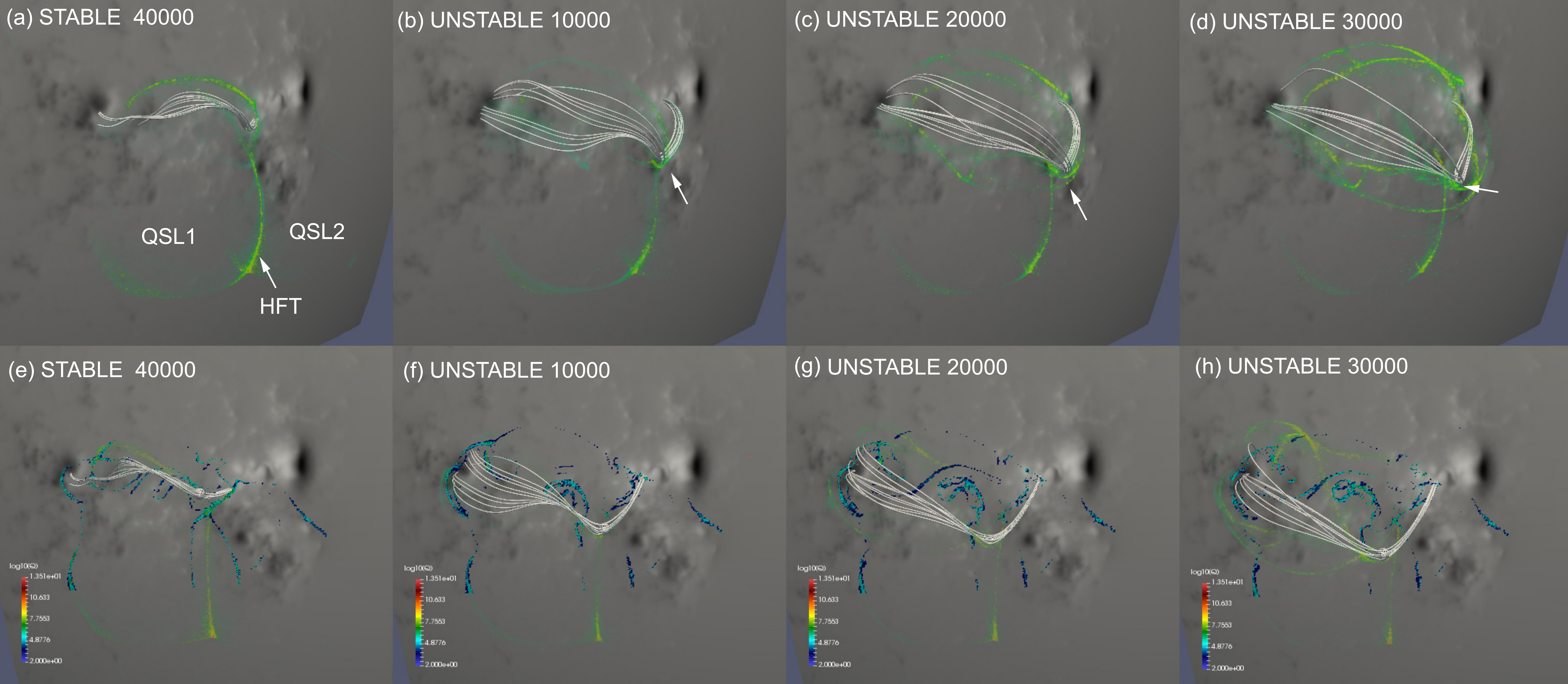}
\caption{Overview of the stable and unstable MFR models with selected field lines from two viewpoints (top row panels: Earth view, bottom row: top view). High log$_{10}$Q-factor values (volumetric rendering in 3D) show the two intersecting QSLs (QSL1 and QSL2) creating the hyperbolic flux tube (HFT) oriented along the N-S. Close to the surface the QSLs roughly correspond to the locations of the flare ribbons (bottom panels (e-h)). The first column (panels (a) and (e)) shows the stable model after 40,000 iterations where it converged. The rest of the columns show the unstable modelled MFR at three snapshots during its evolution (10,000, 20,000, and 30,000 iterations respectively). The selected field lines show MFR field lines that exhibit the ``cusp'' morphology. Note that as the time proceeds the ``cusp'' fieldlines move southward as in the observations.
}
\label{MODEL_SNAPSHOTS}
\end{figure*}

In Figure~\ref{MODEL_SNAPSHOTS} we present renderings of the HFT and MFR lines for the stable and unstable MFR model snapshots. When more axial flux is included to the MFR to make it unstable, the MFR responds by lifting up, but it does not penetrate through the nearby overlying QSL domes. There are clear model field lines that are bent and curved in the same way as in the observations of the cusp-shaped loops. The southward motion of the south cusps is captured in panels (b-d) of Figure~\ref{MODEL_SNAPSHOTS}. However, the model is not able to capture the rapid dynamics and the full evolution of the event, namely the ``cusp splitting''. 

There are two possibilities for this. The first is the sparsity of the set of snapshot cubes. Our first snapshot cube is at 10,000 iterations which may be too late to capture the first set of cusped loops. Numerical diffusion and hyper-diffusion may act quickly at the early stages of the MF relaxation. The second possibility is that we may have not waited long enough for the system to relax. It is usually best to allow the MF relaxation to advance for several thousand iterations as there could be many strong discontinuities that can render the modeled field unphysical. This empirical limitation made us to choose 10,000 iterations as our first snapshot. Nevertheless, the model does portray the essence of the MFR interaction with the topological features (HFT).

The way the relaxation of the unstable MFR proceeds as depicted in the snapshots of Figure~\ref{MODEL_SNAPSHOTS} shows that reconnection occurs at the locations of topologically-induced currents (primarily at the HFT). This reconnection effectively removes the excess axial or poloidal flux (depending on which part of the MFR is in contact with the HFT), which initially caused the MFR to expand. By removing the excess axial or poloidal flux, the MFR weakens and fails to erupt. This is in accordance with the observations, where the filament (cool and dense chromospheric material presumably sitting on the convex-up lines of the MFR structure) does not lift from the PIL throughout the evolution of the phenomenon. This suggests that the removal of the poloidal or axial flux from the HFT/QSLs has brought the, now weakened, MFR back into the stable regime for the ideal MHD Torus instability. In other words, the overlying topology \emph{``shreds''} the unstable MFR and whatever is left of it stays into a stable configuration (switching off the eruption). 

\begin{figure*}
\epsscale{1.0}
\plotone{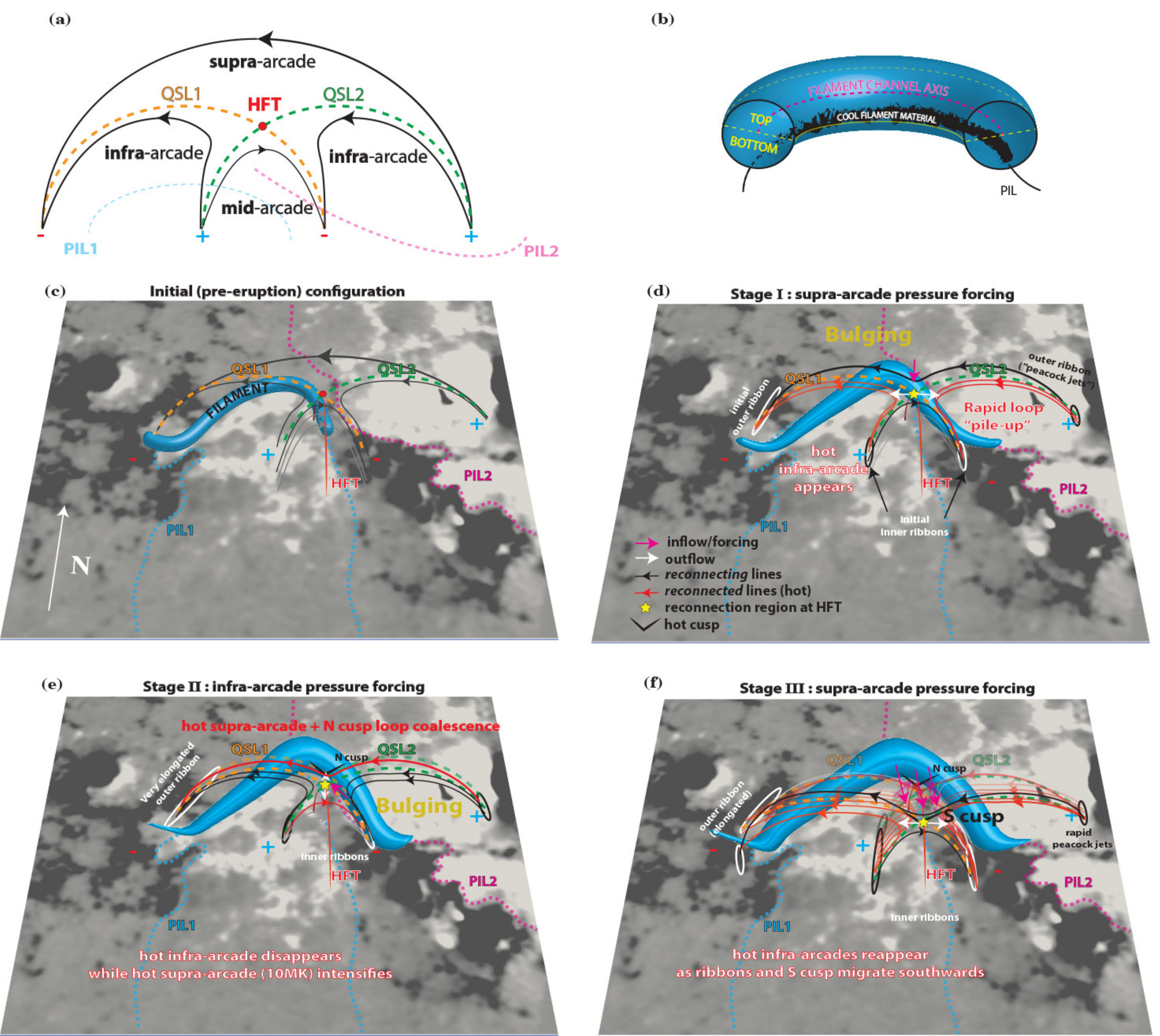}
\caption{(a) Schematic showing the simplified magnetic topology at a vertical cross-section aligned with the E-W direction. The QSL domes and the HFT extend both outwards and inwards the plane of the page. The quadrupolar topology divides the corona into four flux connectivity domains: a ``supra-arcade'' above the QSLs and two ``infra-arcades'' under the QSLs separated by a smaller ``mid-arcade''. (b) Simplified cartoon model of a filament channel. The cool filament material sits on the convex-up lines of a MFR-like structure, the filament channel. These lines correspond to the bottom part of the channel. (c) Simplified 3D graphic representation of the topology shown in panel (a). The filament channel lies below the East infra-arcade. (d-f) The three main stages of the failed eruption. }
\label{CARTOON}
\end{figure*}

In Figure~\ref{CARTOON} we present a synthesis of all the phenomenology into a cartoon model. We have shown that due to the quadrupolar configuration of our target AR we have two QSLs (QSL1 and QSL2) and their intersection results into an HFT. The two intersecting QSLs naturally split the coronal volume into four discrete connectivity domains; a supra-arcade, i.e. the large-scale connectivity of the two exterior polarities outside the QSLs, one infra-arcade per QSL and a mid-arcade in the domain below the intersection of the QSLs (Figure~\ref{CARTOON} (a)). The quadrupolar AR has two PILs under East-West connectivities that support cool filament material and we consider the case of magnetic flux rope-like filament channel. An ideal MHD instability like the torus instability would initially cause the top part of filament channels (Figure~\ref{CARTOON} (b)) to bulge while the bottom part remains relatively inert (and so is the cool filament material). We will now discuss the phenomenology described in Stages I, II, and III summarized in Table~\ref{TABLE} and explain the underlying physical mechanisms behind this failed eruption (see Figure~\ref{CARTOON}).

During Stage I, the filament channel in PIL1 undergoes bulging in its North bend but the PIL1 section underneath QSL1 next to the HFT does not support favorable field line orientation to interact with the HFT. Interaction is possible with the section of the filament channel across PIL2 once it begins to bulge (Figure~\ref{CARTOON}, black field line originating in the positive polarity West of PIL2 and pointing East). The sections of the filament channel that undergo bulging are right at the North of the HFT's northward extent, therefore belonging to the supra-arcade flux domain (outside of the QSLs; Fig~\ref{CARTOON} a). The bulging essentially increases the magnetic pressure locally in the supra-arcade domain above the HFT and forces appropriately oriented MFR field lines to reconnect at the HFT (yellow star, Figure~\ref{CARTOON} (d)) with lines of the mid-arcade and increase the flux content in the infra-arcades (outflow). This is manifested with infra-arcade structures of hot plasma emission in the newly added flux content in these domains, i.e. the hook-shaped loop arcade in Figure~\ref{IRIS_SOT_AIA_evolution} (b) and Figure~\ref{Arcade_evolution_cusps} (a). In addition, ribbons demarkate the footpoints of loops associated with the reconnection event. The rapid loop ``pile-up'' of 400\,km s$^{-1}$ we observed in 131\,\AA\ at the West arcade can be understood to follow the reconnection of MFR fieldlines forced against the HFT due to the bulging MFR. We summarize the phenomenology of Stage I (Table~\ref{TABLE}) as \emph{supra-arcade pressure forcing} leading to \emph{infra-arcade heating}.

In Stage II, the bulging of the filament channel is seen to propagate westward (PIL2; Figure~\ref{Arcade_evolution_cusps} (d-f)) well past the HFT. As the hot West infra-arcade structures suggest, the bulging occurs right next (or even under) QSL2 (Figure~\ref{CARTOON} (e)). This bulging exherts force to the West infra-arcade due to the increased magnetic pressure in this infra-arcade. In turn, this forces flux from the West infra-arcade to reconnect with flux in the East infra-arcade at the HFT. This has the effect of ``killing'' the 6.4 MK hook-shaped East arcade via reconnection with the West infra-arcade and transferring flux to the supra-arcade (East-West large-scale connectivity). The disappearance of the hot infra-arcade in 94\,\AA\ leaves behind a ``pinch''-like gap (Figure~\ref{Arcade_evolution_cusps} (d)) and coincides with enhanced emission seen in 131\,\AA, as an outflow ($\sim$ 250 to 150\,km s$^{-1}$) towards a newly formed North cusp (white arrow pointing North; Figure~\ref{CARTOON} (e)). This succession of events suggests that reconnection between the two hot infra-arcades occurs. We dub this process \emph{infra-arcade pressure forcing}, which leads to \emph{supra-arcade heating}. Note that the North cusp is static (i.e. no proper motions other than coalescence/outflows), which can be understood since it forms right at the North edge of the infra-arcades (primarily due to the forcing exherted by the bulging MFR above PIL2).

In Stage III (which basically has an overlap with Stage II in terms of evolution) we consider the South cusp, which is a dynamic feature evolving quasi-statically (initially moves southwards with 50\,km s$^{-1}$). Following its wake, emission enhances primarily in 94\,\AA, which in the East arcade coincides with the reappearance of the hook-shape arcades along the southward direction. We interpret this observation again as \emph{supra-arcade pressure forcing} (i.e. forcing of loops to reconnect from above the HFT) for two reasons. First, the loop coalescence towards the concave part of the South cusp is only seen in 131\,\AA, which clearly shows the large-scale nature of the connectivity following the South cusp. Second, the loop coalescence is comprised of loops traversing clearly 20$\arcsec$ from N to S, a large distance over which the loops are tracked and resolved (Figure~\ref{Cusp_panels_and_slitstackplot} (c)). In addition to these reasons, our MF model for an unstable filament channel reveals an almost identical evolution for the South cusp. That is, an expanding (i.e. bulging) top part of the unstable model MFR creates a supra-arcade ``front'' at its intersection with the HFT that propagates southward. At each time, this intersection of MFR supra-arcade lines with the HFT manifests itself as a moving front $-$ the South cusp.

\begin{figure*}
\epsscale{1.0}
\plotone{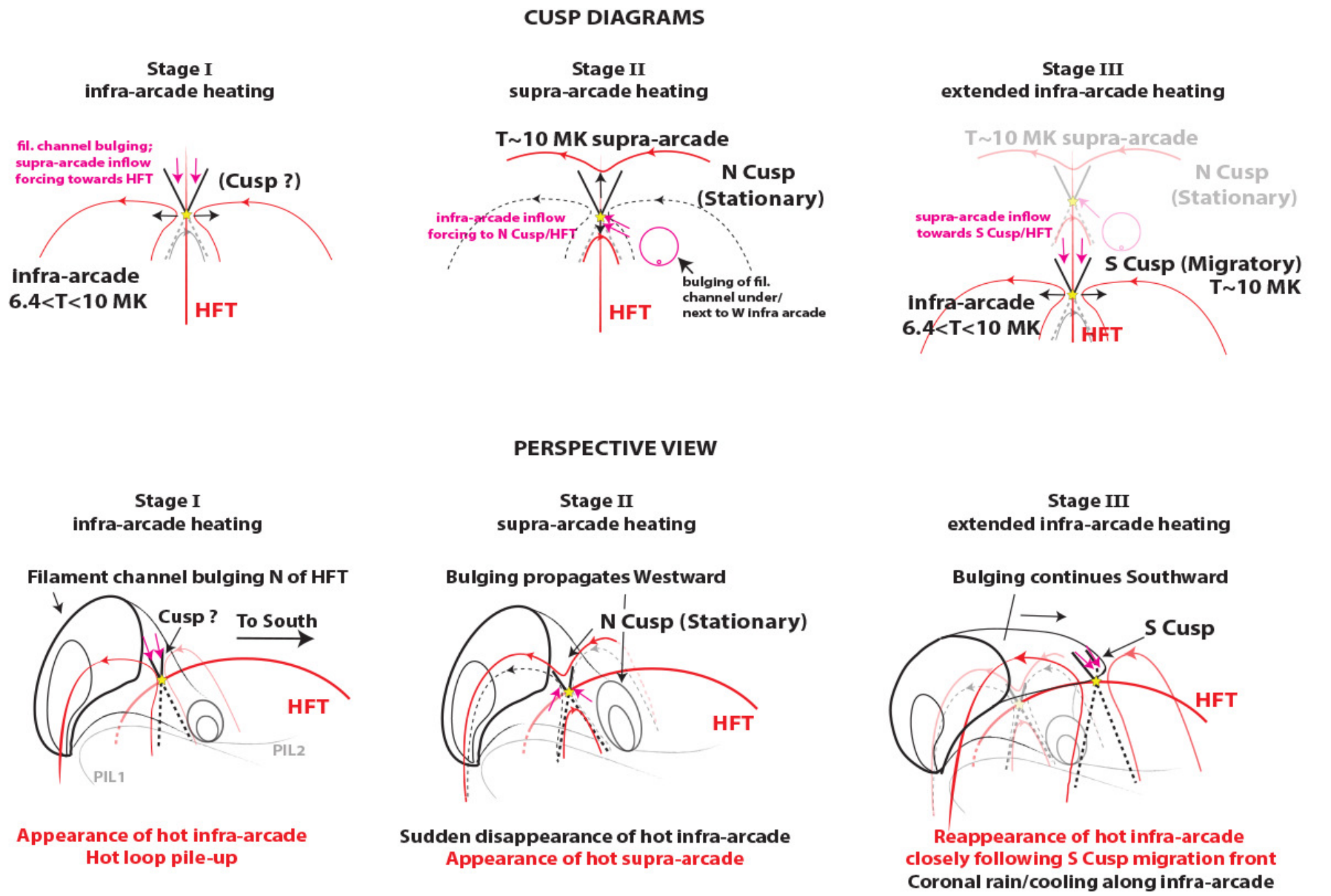}
\caption{Top row: Simplified cusp diagrams explaining the phenomenology of the failed eruption. Bottom row: Simplified perspective view to illustrate the forcing of field lines due to the bulging of the filament channel in the infra and supra arcades. Magenta arrows denote inflows/forcing of field lines towards the HFT. Dashed black loops denote disappearance after their reconnection. Newly reconnected field lines are shown with red color.}
\label{CUSP_DIAGRAM}
\end{figure*}

We therefore conclude that the expanding MFR causes \emph{supra-arcade pressure forcing}, which further consumes flux from the expanding MFR through reconnection at the HFT, inflow speeds of -190 to -150\,km s$^{-1}$ and re-heating of the infra-arcades (Figure~\ref{CARTOON} (f)). Both Stage I and Stage III are interpretted as \emph{supra-arcade pressure forcing} for the same reason (expanding top of an MFR forcing reconnection). The succession of events suggests that Stage III is a natural continuation of Stage I and, in essence, they are associated with the initial and later stages of MFR expansion reconnecting at the HFT. A more simplified diagram focusing on the nature of the cusps is in Figure~\ref{CUSP_DIAGRAM}.

 \citet{Liu_R_etal_2014} reported a failed eruption with similarities to our event; namely, a confined flare, quasi-static cusp structures and ribbon brightenings. The authors used an NLFFF model and deduced that the event  involved two adjacent sheared arcades (one of them containing an MFR) separated by a ``T-type'' HFT. They concluded that a flux emergence event triggered the flare, and the HFT and the MFR dictated the structure and dynamics of the flare loops. They also mention the existence of a QSL high above the HFT as an attempt to explain the cusp shaped loops. The loop coalescence reported in \citet{Liu_R_etal_2014} happens at a much slower rate - they mention ``a few tens of km\,s$^{-1}$'' - much lower than our reported speeds of $>$150 km\,s$^{-1}$. In addition, the confined flare is quite strong, i.e. an X1.9 flare. The event in their case study is rather energetic compared to the one in this paper. In their view, the reconnection at the HFT explained the long duration of the X1.9 flare, while the flare peak happens at the Bald Patch Separatrix Surface (BPSS) below the MFR (i.e. the MFR lifts but stays confined). In our observations the flare is of a rather low magnitude, a mere C2.0. Since the MFR never fully lifts from the surface (filament material sits stably above the PIL), this means that the activity we observe is predominantly due to the reconnection of the top parts of the MFR at the HFT and not due to reconnection below the MFR.

\section{Conclusions}

Thanks to the coordinated campaign in support of the \textit{VAULT2.0} rocket launch, we were able to capture the initial stages of a failed eruption with unprecedented temporal, spatial and spectral coverage. In contrast to the majority of modeling and observational work on the role of magnetic topology in intiating CMEs, our analysis shows, for the first time, how magnetic topology can suppress ejections already in progress. In a nutshell, a Hyperbolic Flux Tube can shred a rising flux rope attempting to pass through it and results in a failed eruption. We summarize the main points of our analysis as follows:

\begin{enumerate}
\item Observed heating along the intersection of the two Quasi-Separatrix Layers, i.e. heating along the Hyperbolic Flux Tube (seen in 131\,\AA; matches with topology recovered from extrapolation). 

\item Observed cooling downflows along filament (seen by \textit{VAULT2.0}, \textit{IRIS}, SOT). 

\item The ribbon-like simultaneous brightenings at remote locations during the activity suggest the presence of topological structures (Quasi-Separatrix Layers, Hyperbolic Flux Tube).

\item No CME.

\end{enumerate}

The conclusion of our analysis can be summarized as follows: 

\begin{enumerate}
\item The expanding magnetic flux rope meets the nearby Quasi-Separatrix Layers \& Hyperbolic Flux Tube, shreds (reconnects) and cool material returns back to the lower atmosphere as coronal rain.

\item Reconnection during Stage I and Stage III (supra-arcade forcing) essentially transfers flux from the flux rope and repartitions it to the subdomains under the Quasi-Separatrix Layers (infra-arcades).

\item Lack of filament rise means that the flux rope shreds and weakens before its rise phase kicks in and before subsequent ``flare reconnection'' could take over and eject the filament/flux rope as a CME.

\end{enumerate}

Thus, topology is not only acting to facilitate an eruption as it has been previously reported in the literature, but it may also lead to the ``killing'' of an unstable magnetic flux rope, \emph{preventing an eruption to occur}. This work presents the benefits of observations in high time-cadence (first observation of eruption initiation in high cadence from SOT \ion{Ca}{2}; 3\,s) and excellent temperature coverage from the campaign observations in support of the \textit{VAULT2.0} mission.

\acknowledgments
The authors are grateful to the anonymous referee for the careful review and constructive comments. The authors wish to thank Drs. V. Titov, B. Kliem, T. T\"{o}r\"{o}k and the International Space Science Institute (Bern, Switzerland) for supporting the International Working Team on Decoding the Pre-Eruptive Magnetic Configuration of Coronal Mass Ejections led by S. Patsourakos and A. Vourlidas for valuable discussions. G. Chintzoglou also thanks Prof. C.E. Alissandrakis, A. Nindos, J. Zhang and J. Karpen for inspiring conversation and encouragement. G.C. acknowledges support by NASA contract NNG04EA00C (SDO/AIA). G.C. was also supported by the NASA Earth and Space Science Fellowship Program - Grant NNX12AL73H. G.C. thanks NASA and the Wallops Space Flight Center for their approval to participate as a graduate student at the \emph{VAULT2.0} launch operations. The \textit{VAULT2.0} project and launch operations (A.V., G.C., S.T.B.) were supported by NASA NNG12WF67I. The participation of A.V in this work was partially funded by the NASA LWS program through ROSES NNH13ZDA001N. A.S. and S.T. acknowledge partial support by NASA contract NNM07AB07C to SAO, support by the Air Force Office of Scientific Research under award FA9550-15-1-0030 to UCAR and subaward to SAO Z15-12504 AFGL and NASA HSR grant to SAO NNX16AH87G. The work of G.S. was supported by NASA contract S-13631-Y to NRL. {\it Hinode} is a Japanese mission developed, launched, and operated by ISAS/JAXA in partnership with NAOJ, NASA, and STFC (UK). Additional operational support is provided by European Space Agency (ESA), NSC (Norway). \emph{IRIS} is a NASA small explorer mission developed and operated by LMSAL with mission operations executed at NASA Ames Research center and major contributions to downlink communications funded by ESA and the NSC. HMI and AIA are instruments on board \emph{SDO}, a mission for NASA's Living with a Star program.


\end{document}